\documentclass[twocolumn,prb,aps,superscriptaddress]{revtex4}
\usepackage{color}
\usepackage{graphicx}
\usepackage{natbib}
\newcommand{\nc}{\newcommand}
\nc{\on}{\operatorname}
\nc{\wt}{\widetilde}
\nc{\Wick}{{\mathbb :}}
\nc{\R}{{\mathbb R}}

\newcommand{\beq}{\begin{equation}}
\newcommand{\eeq}{\end{equation}}
\newcommand{\bmul}{\begin{multline}}
\newcommand{\emul}{{\end{multline}}}
\newcommand\beqa{\begin{eqnarray}}
\newcommand\eeqa{\end{eqnarray}}
\newcommand\bea{\begin{array}}
\newcommand\eea{\end{array}}
\newcommand\ba{\begin{array}}
\newcommand\ea{\end{array}}
\newcommand{\nn}{\nonumber}

\newcommand{\neqa}{\nonumber\end{eqnarray}}

\newcommand{\eq}[1]{eq.(\ref{#1})}
\newcommand{\eqs}[2]{eqs.(\ref{#1},\ref{#2})}

\renewcommand{\d}{\partial}

\nc{\CH}{{\mathcal H}}
\nc{\Db}{{\bar D}}
\nc\comment[1]{}

\nc{\CM}{{\mathcal M}}
\nc{\CN}{{\mathcal N}}

\newcommand{\re}{\relax{\rm I\kern-.18em R}}

\renewcommand{\Im}{{\mathrm {Im}}}
\renewcommand{\Re}{{\mathrm {Re}}}
\nc{\meV}{{\mathrm{\,meV}}}
\nc{\cG}{{\mathcal G}}

\renewcommand{\)}{\right)}
\renewcommand{\(}{ \left( }
\comment{ \)}
\renewcommand{\bar}{\overline}

\nc{\al}{{\alpha}}

\def\eps{{\epsilon}}
\def\eV{{\rm eV}}
\def\cO{{\cal O}}

\begin{document}

\title{Magnetic fluctuations and specific heat in Na$_x$CoO$_2$ near a
Lifshitz  Fermi surface  topological transition}
\author{Sergey Slizovskiy}
\affiliation{Department of Physics, Loughborough University, Loughborough LE11 3TU, UK}
\author{Andrey V. Chubukov}
\affiliation{Department of Physics,University of Wisconsin-Madison, Madison, WI 53706, USA}
\author{Joseph J. Betouras}
\affiliation{Department of Physics, Loughborough University, Loughborough LE11 3TU, UK}
\begin{abstract}
We analyze the temperature and doping dependence of the specific heat $C(T)$ in Na$_x$CoO$_2$.
This material
 was
conjectured to
undergo a Lifshitz -type topological transition at $x =x_c =0.62$, in which a new  electron Fermi pocket emerges at the $\Gamma$ point, in
addition to the
existing hole pocket with large $k_F$. The data
show that near $x =x_c$, the temperature dependence  of $C(T)/T$  at low $T$ gets stronger as $x$ approaches $x_c$ from below and
then reverses the trend and changes sign at $x \geq x_c$. We argue that this behavior can be quantitatively explained within the
spin-fluctuation
theory. We show that magnetic fluctuations are enhanced near $x_c$ at momenta around $k_F$
  and their dynamics changes between  $x \leq x_c$ and
 $x >x_c$, when the new pocket forms. We demonstrate that this explains the temperature dependence of $C(T)/T$.
 We show that at larger $x$ ($x > 0.65$)
the system enters a magnetic quantum critical regime where $C(T)/T$ roughly scales as $\log T$. This behavior extends to progressively
lower $T$
 as $x$ increases towards a magnetic instability at $x \approx 0.75$.
\end{abstract}
\maketitle

\begin{figure}
\includegraphics[scale=0.4]{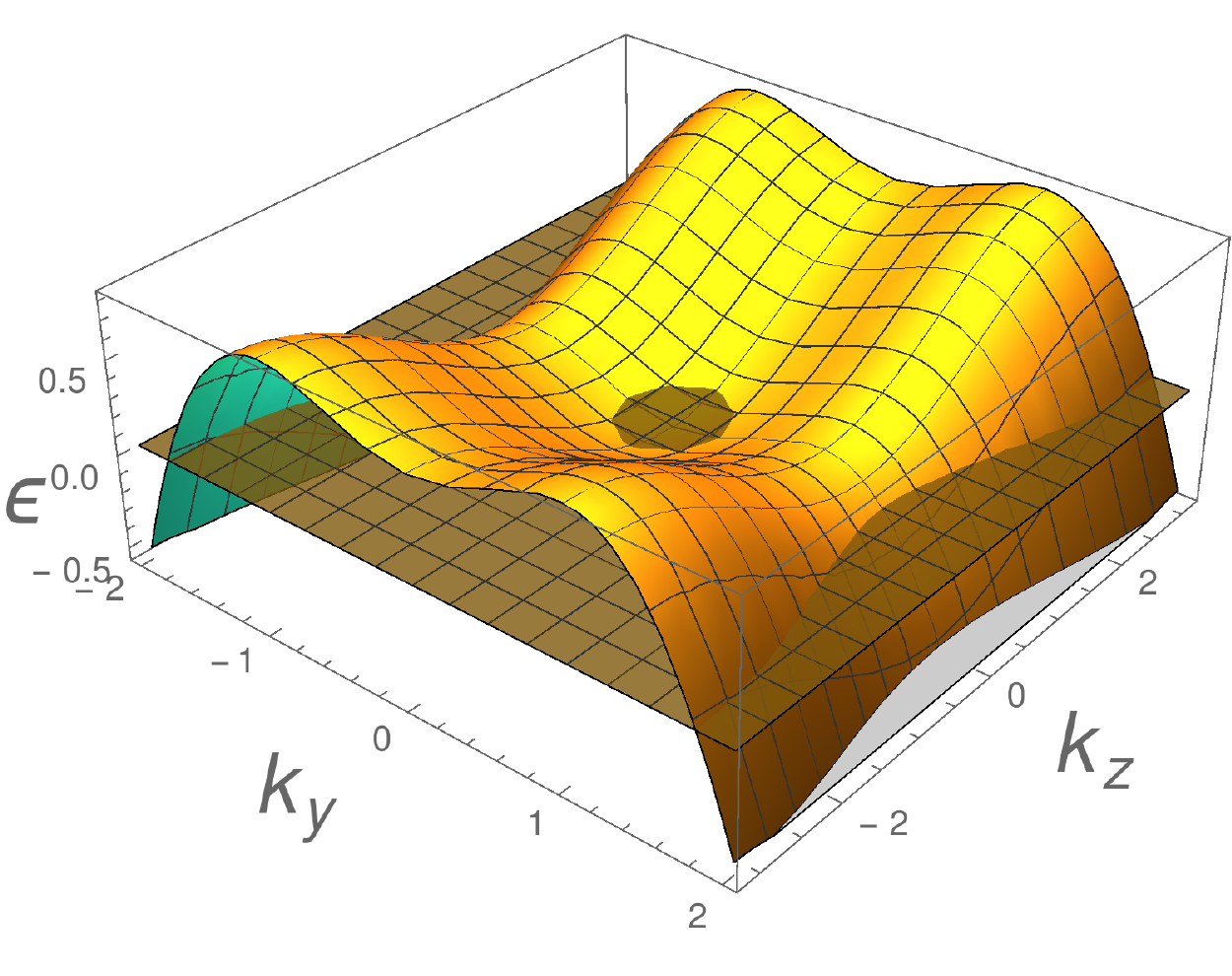}
\caption{\label{fig:Dispersion} \label{fig:Chi0}
The lattice fermionic dispersion $\epsilon (k)$ at $k_x = 0$ (in units of $t_1 \approx 0.1 eV$).
See~\cite{disp} for the values of the other hopping integrals. Note that the dispersion is approximately rotationally invariant in the
$k_x - k_y$ plane
and is quite shallow: the depth of the local minimum is around $20\meV$.}
\end{figure}

{\it  Introduction}~~~The layered cobaltates  Na$_x$CoO$_2$ have been the subject of intense studies in recent years due to their very
rich phase diagram
and associated rich physics \cite{Wang, Lee, Balicas, Li, Singh, Johannes, Helme}. Their structure is similar to that of copper oxides and
consists of
alternatively stacked layers of CoO$_2$ separated by sodium ions. The Co atoms form a triangular lattice \cite{Fouassire}.  The hydrated
compound
Na$_x$CoO$_2$:yH$_2$O with $x \sim 0.3$ shows superconductivity \cite{Takada}, most likely of electronic origin. The anhydrated parent
compound
Na$_x$CoO$_2$ exhibits low resistivity and thermal conductivity and high thermopower~\cite{Wang, Lee} for $0.5 < x <0.9$ and magnetic
order for
$0.75<x<0.9$~(Refs.\onlinecite{Johannes,Helme,Motohashi,Bayrakci}).  In the paramagnetic phase Na$_x$CoO$_2$ shows a conventional metallic
behavior at
$x \leq 0.6$ and at larger $x$ displays strong temperature dependence of both spin susceptibility and specific heat down to very low $T$ .
This change of behavior has been attributed~\cite{Okamoto} to a putative Lifshitz-type topological transition \cite{Lifshitz} (LTT) at
$x_c \approx 0.62$, in which a small three-dimensional (3D) electron Fermi pocket appears around $k=0$, in addition to the already
existing
quasi-2D hole pocket with large $k_{F1}$ (Ref.\onlinecite{Korshunov}), see Fig. \ref{fig:Dispersion}.
Although the small pocket has not yet been observed directly, ARPES measurements at smaller $x$ did find a local
minimum in the quasiparticle dispersion at the $\Gamma$ point \cite{Arakane}.
Similar topological transitions have been either observed or proposed for several solid state
[\onlinecite{Katsnelson, Hackl, Chen, LeeStrackSachdev, ChubukovMorr, Liu, Zwicknagl, Porret}] and cold atom systems [\onlinecite{QCB}],
and the understanding of the role played by the interactions near the LTT transition is of
rather general interest to condensed matter and cold atoms communities.

\begin{figure*}[htp]
$\begin{array}{ccc}
\includegraphics[scale=0.41]{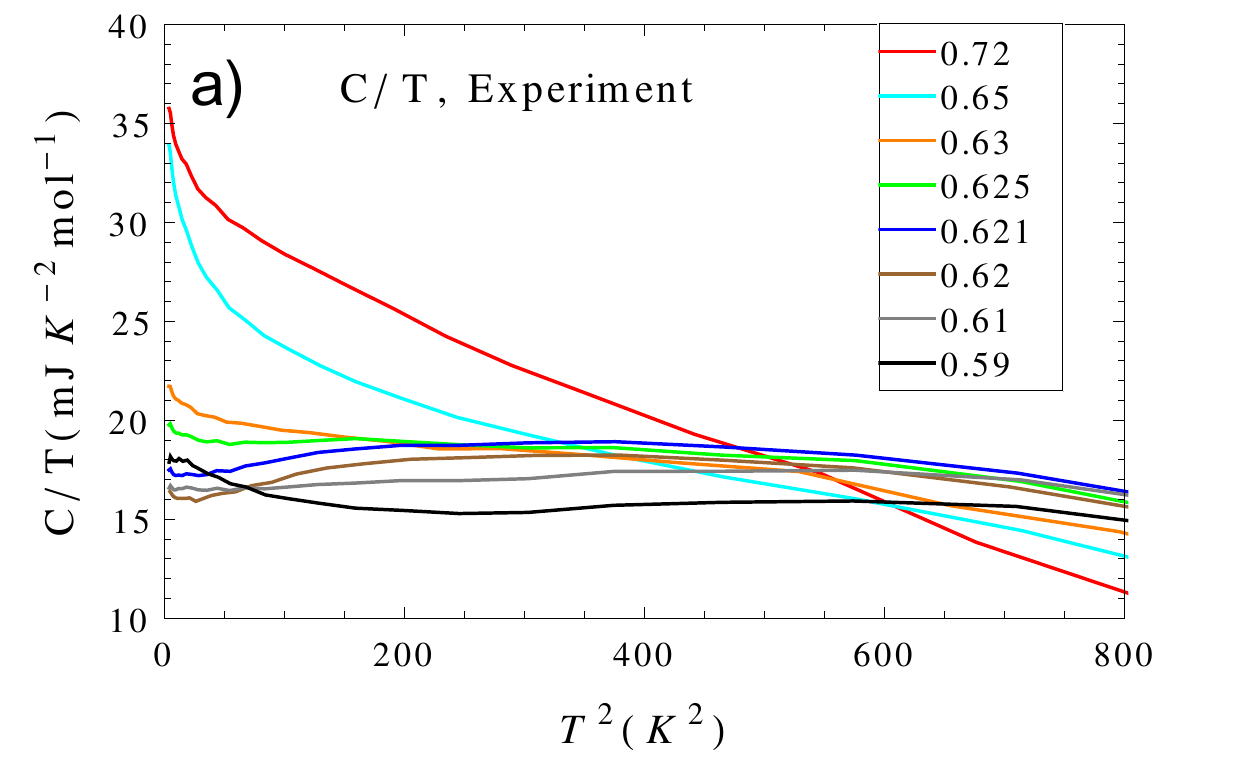} &
\includegraphics[scale=0.44]{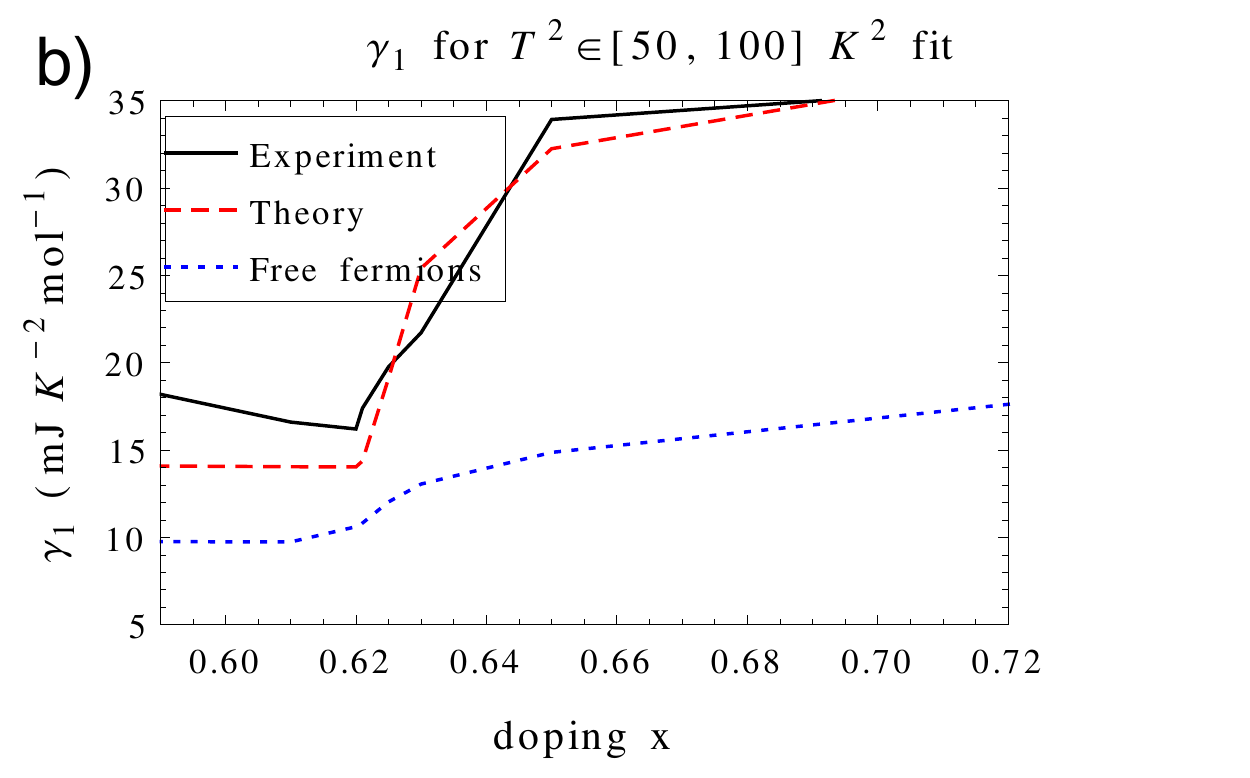}
\includegraphics[scale=0.43]{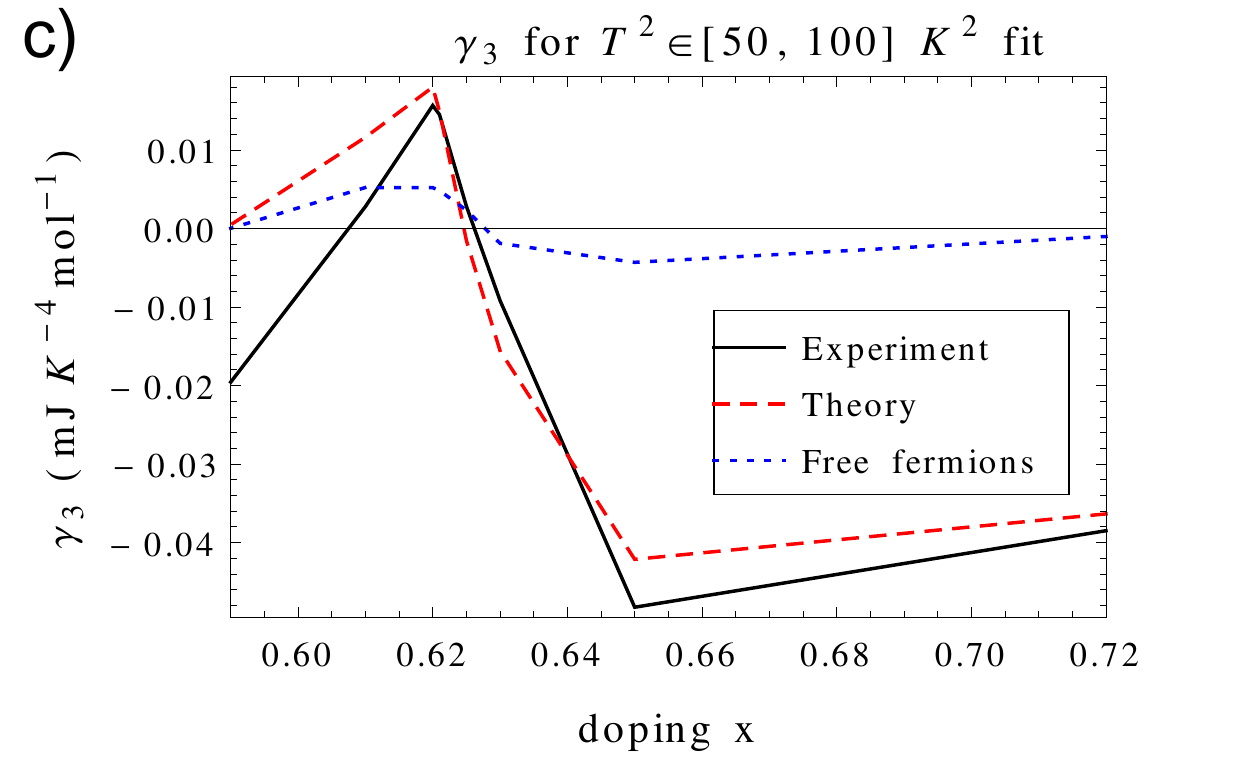}&
\end{array}$
\vspace{-0.5cm}
 \caption{\label{fig:CvFree_2}
(a) The data~\cite{Okamoto}  for $C(T)/T$ for $x=0.59$ to $0.72$  with the doping-independent phonon contribution
subtracted.
 (b,c) The fits of experimental and theoretical
$C(T)/T$ to $C(T)/T = \gamma_1 + \gamma_3 T^2$ for
$T^2$ between $50K^2$ and $100K^2$.}
\vspace{-0.5cm}
\end{figure*}

\begin{figure*}[htp]
$\begin{array}{cc}
\includegraphics[scale=0.55]{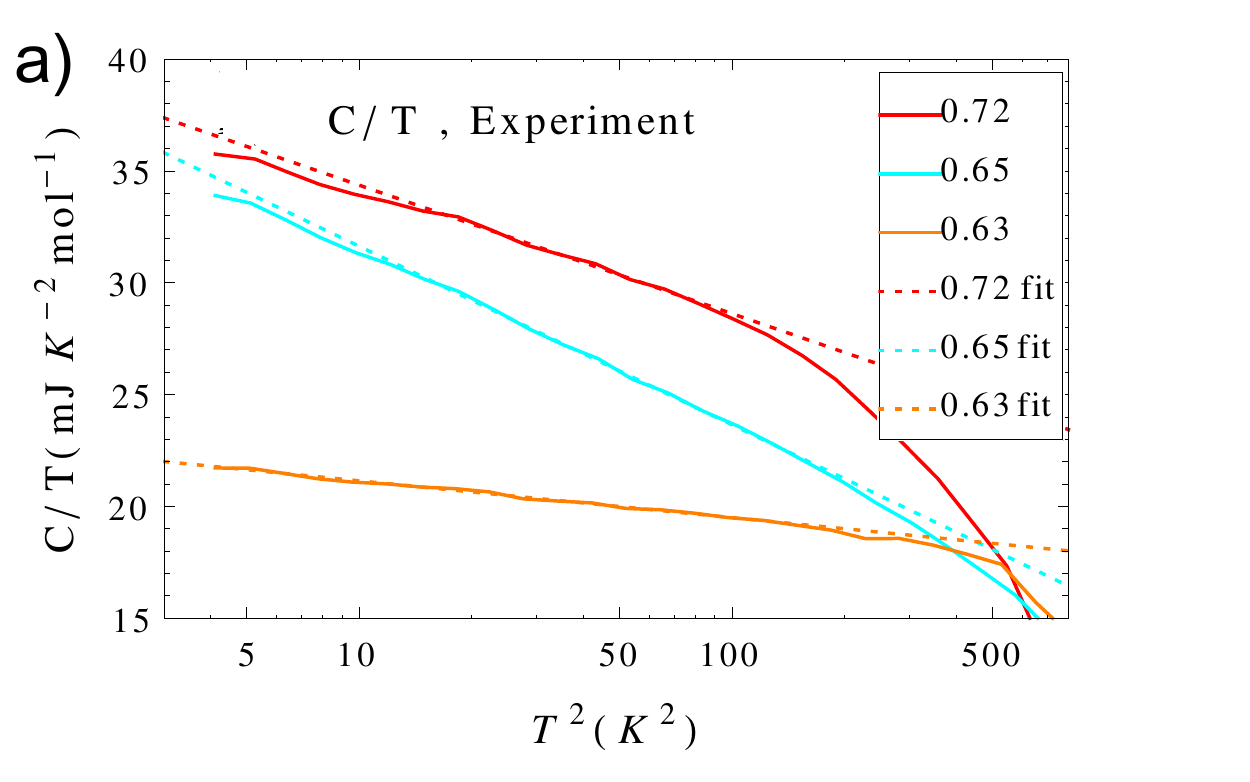}&
 \includegraphics[scale=0.55]{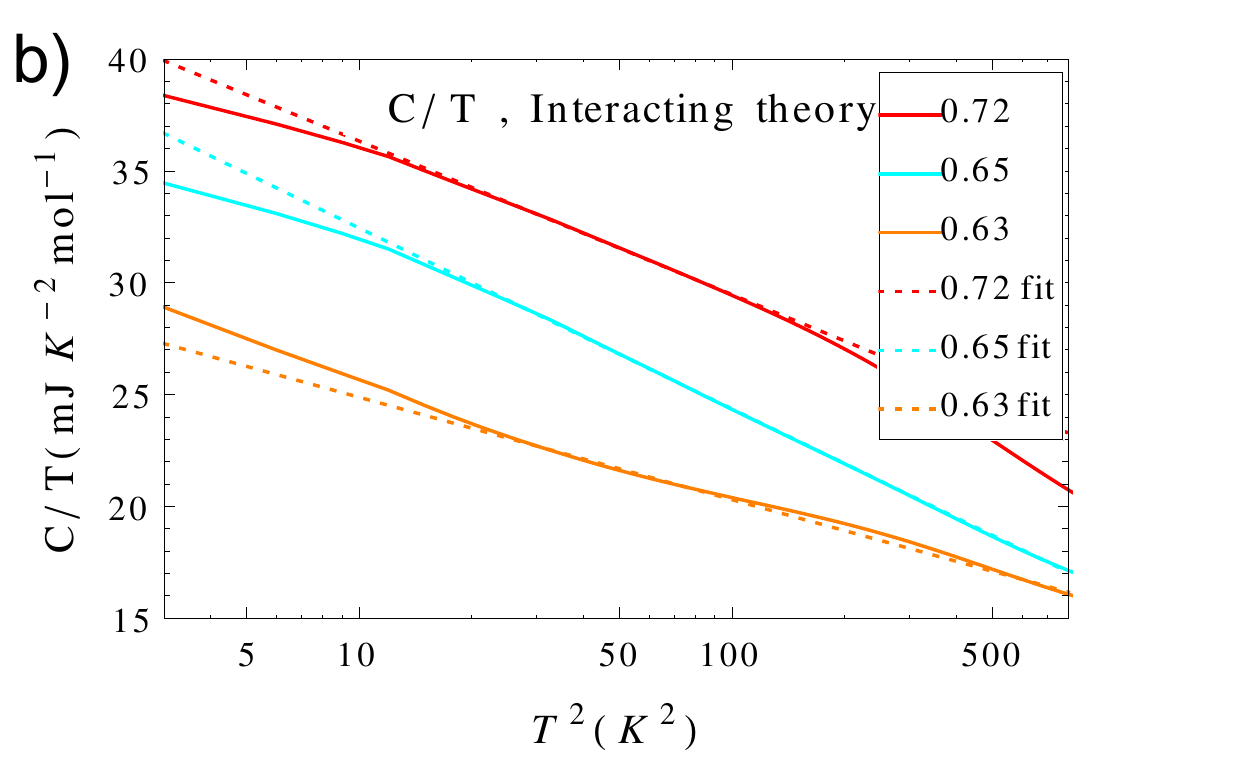}
\end{array}$
\vskip-0.5cm
 \caption{\label{fig:CvFree_3}
 Experimental data for doping $x=0.63, 0.65, 0.72$ from Ref.\onlinecite{Okamoto} (a) and theoretical (spin-fluctuation) result (b) for
$C(T)/T$ in semi-logarithmic temperature scale.
The dashed lines correspond to $C(T)/T \propto \log T$ fit.
The prefactor of the $\log T$ depends on magnetic correlation length $\xi$
 }
\end{figure*}

The subject of this paper is the analysis of interaction contributions to the specific heat $C(T)$ in Na$_x$CoO$_2$ at around the critical
$x_c$ for LTT.
The experimental data \cite{Okamoto},
  show (see Figs. \ref{fig:CvFree_2} and  \ref{fig:CvFree_3})
   that for doping near $x_c$, the temperature dependence of $C(T)/T$ is more complex than the
$C(T)/T = \gamma_1 + \gamma_3 T^2 + O(T^4)$ expected in an ordinary Fermi liquid (FL). The FL behavior itself is not broken in the sense
that $\gamma_1$ remains finite.
However the $T$ dependence  at $x = x_c$ is stronger than $T^2$, as evidenced by the fact that the fits of the data on $C(T)/T$  to
$\gamma_1 + \gamma_3 T^2$ behavior~\cite{Okamoto} in finite intervals around different $T$ yield larger $\gamma_3$
 as $T$ goes down
 (see Ref.\onlinecite{SI}).
This does not allow one to interpret $\gamma_1$ directly as a density of states, and the full computation is needed to compare the data 
with
the theory.
 For  doping levels $0.65 < x <0.75$  the data show~\cite{Balicas}
 that,  to a good approximation, $C(T)/T \propto \log T$ in a wide range of temperatures $T \sim$ 1 -- 10 K, see Fig.
\ref{fig:CvFree_3}a.
This logarithmic temperature dependence progressively spans over larger temperature range as $x$ approaches 0.75,
 where a magnetic order develops (Refs.[\onlinecite{Johannes,Helme,Motohashi,Bayrakci}]).

\begin{figure*}[htp]
$\begin{array}{cc}
\includegraphics[scale=0.52]{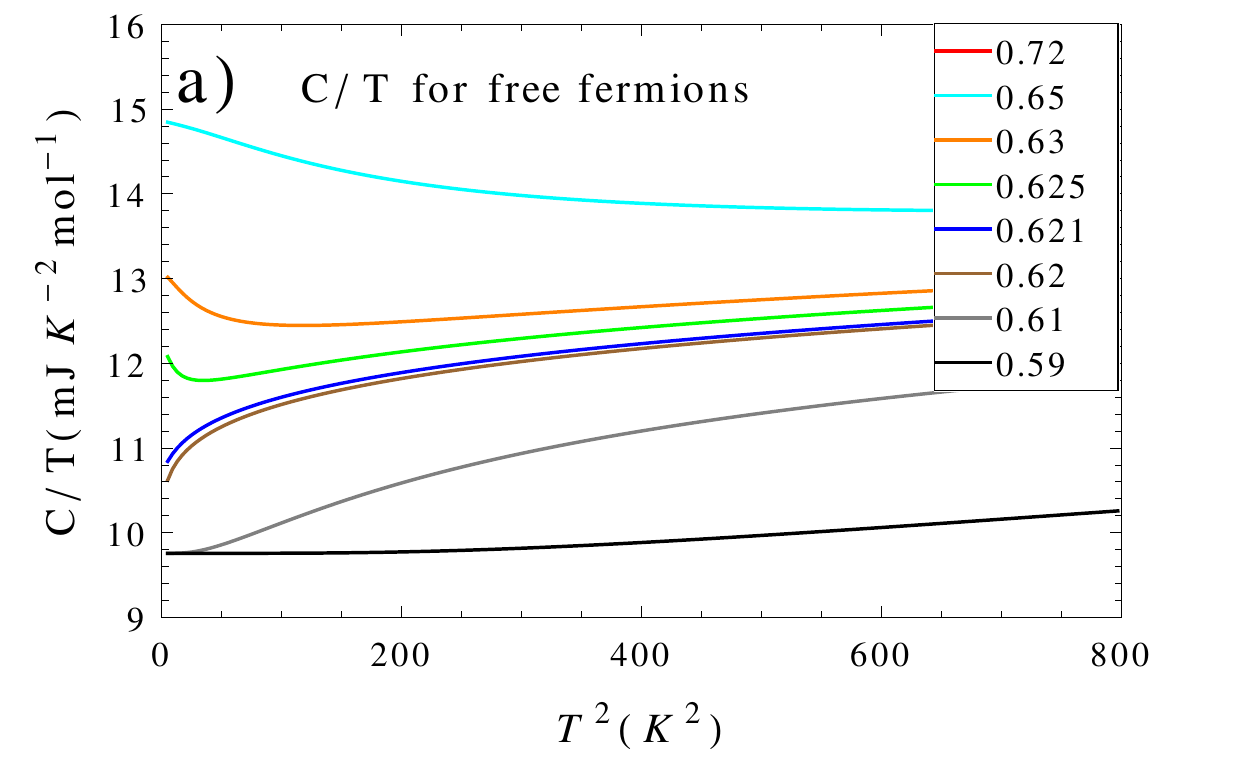} \hspace{-0.3cm}&
\includegraphics[scale=0.5]{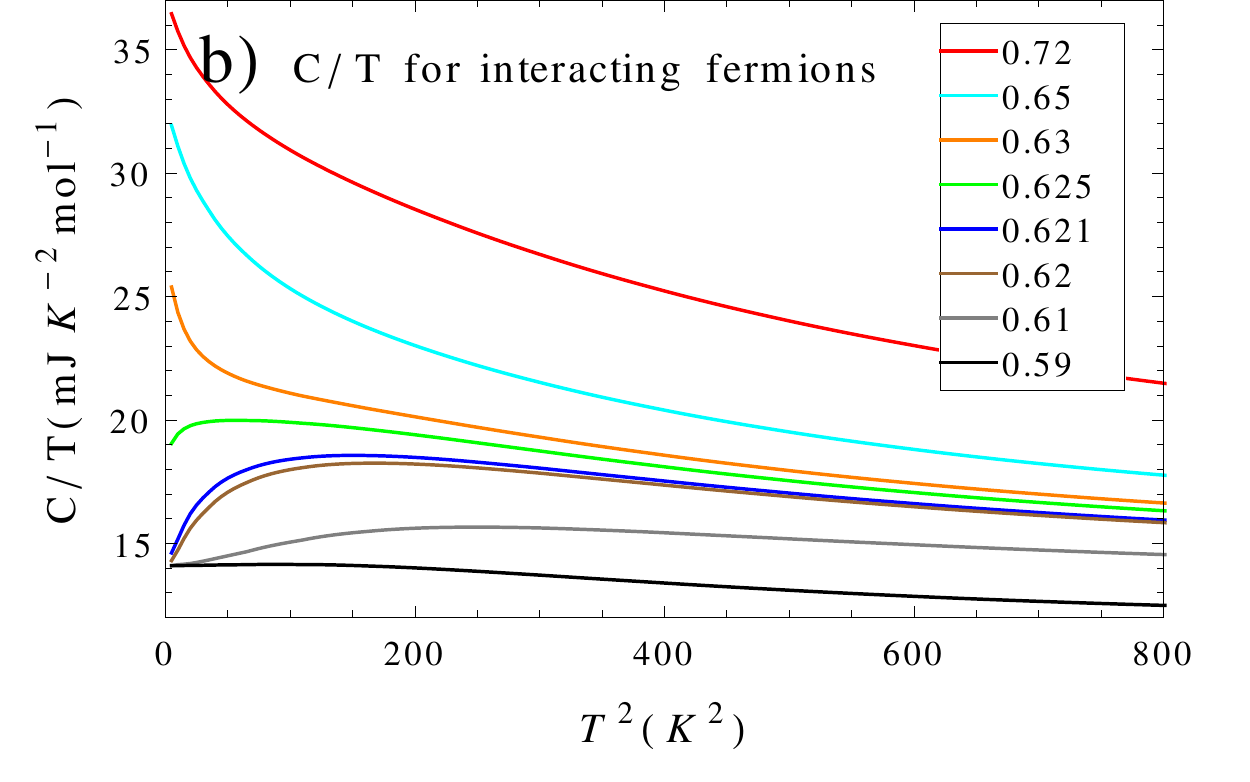} \hspace{-0.5cm}
\end{array}$
\vspace{-0.5cm}
 \caption{\label{fig:CvFree_1}
Theoretical results for the specific heat $C(T)/T$ for for several  Na dopings $x$ for free fermions (a) and for fermions with 
magnetically-mediated
interaction with $\xi=7 a_0$ (b).
Both are obtained without expanding in $T$, using  the dispersion from Fig.\ref{fig:Dispersion}.
 }
\end{figure*}

Some qualitative features of the experimental data of $C(T)$ at $x \sim x_c$ are reproduced by the free-fermion formula for specific
heat, with the
quasiparticle dispersion taken from first-principle calculations (Fig. \ref{fig:CvFree_1}a).
In particular, $\gamma_1$ increases and $\gamma_3$ passes through a maximum
 around $x =0.62$, see Fig. \ref{fig:CvFree_2}b,c .
However, the magnitudes of $\gamma_1$ and $\gamma_3$ are much smaller than in the data
 and the maximum in $\gamma_3$ is too shallow. A strong temperature dependence of $C(T)/T$ may potentially come from phonons,
but $\gamma_3$ due to phonons is highly unlikely to become singular at $x= x_c$.
 This implies that the observed features of $C(T)$ are most likely caused by electron-electron interactions.
Interactions with a small momentum transfer $q$ give rise to linear in $T$ dependence of $C(T)/T$ in 2D due to non-analyticity associated
with the
Landau damping \cite{Belitz}. That a linear in $T$ term has not been observed in Na$_x$CoO$_2$ near $x_c$ implies that small-$q$
fluctuations are weak near
this doping\cite{comm1}. Interactions with a finite momentum transfer $q \approx k_{F1}$ are expected to be strong and
sensitive to the opening of a new piece of electron FS as the static fermionic polarization operator $\Pi (k_{F1})$ gets enhanced as $x$
approaches $x_c$.
An enhancement of $\Pi (k_{F1})$ generally implies that spin fluctuations at $k_{F1}$ get softer and mediate fermion-fermion interaction
at low
energies \cite{comm1}.

The spin-fluctuation contribution to $\gamma_3$ has been analyzed before for systems with a single 3D FS\cite{millis}. In
this situation, the sign of $\gamma_3$  is negative.
 This  negative sign can be traced
back~\cite{millis} to  positive sign of the prefactor for the $\omega^2$ term in the dynamical spin
susceptibility $\chi (q, \omega)$. The latter behaves at small frequencies and at momenta $q <2k_F$, which connects points on the FS, as
 $\chi^{-1} (q, \omega) \propto \xi^{-2} + b \omega^2 - i \gamma \omega$ with $b \propto 1/q^2 >0$.
We show that
in our case relevant momenta are around $k_{F_1}$ and situation with $b >0$ holds for $x >x_c$, when a small 3D pocket emerges
 and  $k_{F_1}$ connects fermions at the two FSs.  For $x <x_c$, when only 2D FS is present, we found that
  the sign of  $b$ is negative. This gives rise to positive $\gamma_3$ at $x \leq x_c$ and negative $\gamma_3$ at $x >x_c$, consistent 
with the
   data in Na$_x$CoO$_2$ (see Fig. \ref{fig:CvFree_2}b,c).  We further show that $b$ is singular at small $\mu$ and this gives rise to 
non-monotonic behavior of
   $\gamma_3$ around $x_c$ -- it increases upon approaching $x_c$ from below,
   passes through a maximum and then rapidly decreases and changes sign at $x \geq x_c$ (Fig. \ref{fig:CvFree_2}c).
     We argue that this behavior is fully consistent with the data.

 When the temperature exceeds $1/(\xi^2 \gamma)$,
 the system enters into a quantum-critical regime.
We found that in this regime, the specific heat can be well fitted by  $C(T)/T \propto \log T$ (see Fig. \ref{fig:CvFree_3}).
 The lower boundary of quantum-critical
behavior extends to lower $T$ as $x$ increases towards the onset of a magnetic transition at $x \approx 0.75$.
This is again consistent with the experiment~\cite{Balicas} which observed
$C(T)/T \propto \log T$ down to  0.1 K at $x = 0.747$.

{\it  The model.}~~~We follow earlier works\cite{Kuroki2007, Korshunov} and consider fermions with the tight-binding dispersion
$\epsilon (k)$ on a triangular lattice with hopping up to second neighbors in $xy$ plane and to nearest neighbors along
$z$-direction~\cite{disp}.
The dispersion, shown in Fig. \ref{fig:Dispersion}, has a hole-like behavior at large momentum ($\partial \epsilon (k)/\partial k <0$) and
a local
minimum at the $\Gamma$ point ${\bf k} =0$. At $\mu <0$, ($x < x_c=0.62$) the Fermi surface consists of a single quasi-2D hole pocket with
large
$k_F = k_{F1}$. As $\mu$ crosses zero and becomes positive, a new 3D Fermi pocket appears, centered at the $\Gamma$ point (see
Fig. \ref{fig:Dispersion}). For the specific heat analysis at small $|\mu|$ we can approximate the dispersion near $k=0$ by
$\epsilon (k) =k^2/(2 m) + k_z^2/(2 m_z)$ and approximate the large Fermi surface by an effectively 2D dispersion $\eps (k) \approx
v_{F1} (k-k_{F1})$, where $k= \sqrt{k_x^2+k_y^2}$.
 In our analysis, we do not consider $Na$ charge ordering. Such an ordering does indeed develop at intermediate
dopings~\cite{alloulNew,kotliar}.
However, the experimentalists, who performed the 
measurements on quenched samples of Na$_x$CoO$_2$ at $x \geq 0.6$, 
did not observe time-dependent phenomenon and argued~\cite{okamoto_1} that their quenched samples are in quasi-equilibrium state.

{\it $C(T)$ for free fermions.}~~~To set the stage for the analysis of interaction effects we first compute the specific heat for free
fermions with non-monotonic dispersion $\eps (k)$. The grand canonical potential is given by
\beq
 \Omega (T, \mu, V) = -T \int \rho(\eps) \ln(1+e^{-(\eps-\mu)/T}) d\eps,
\eeq
Evaluating the entropy $S (T, \mu, V)$,
extracting $\mu = \mu (T, V)$ from the condition on the number of particles
and expanding $C (T) = C_V (T) = T \left(\frac{\partial S}{\partial T}\right)_{V}$ in temperature,
we obtain
  at the lowest $T$
\beqa \label{cvfree}
&&C(T)/T =\gamma_1  + \gamma_3 T^2 + O\left(T^4\right) \nonumber \\
&&\gamma_1 = \frac{\pi ^2 \rho }{3}, ~~\gamma_3 = \frac{\pi^4}{30} \frac{ \left(7 \rho  \rho ''-5 \left(\rho
'\right)^2 \right)}{\rho}
\eeqa
where $\rho(\mu)$ and its derivatives over $\mu$ are computed at $T=0$. The low-T expansion in (\ref{cvfree}) is valid for $T < |\mu|$.
Analyzing (\ref{cvfree}), we find that for $\mu < 0$, when there is no electron pocket, the $T$ dependence  comes from a large hole
pocket and is  non-singular.
 For $\mu >0$, the electron pocket appears with $\rho (\mu) \propto \sqrt{\mu} \theta(\mu)$. This gives rise to negative $\gamma_3$, which
diverges
at small $\mu$ as $1/{\mu}^{3/2}$. At $\mu = 0$ the analytic expansion in powers of $T^2$ doesn't work even at the lowest $T$. We
found\cite{SI} that in
this case
\beq\label{HighT}
\frac{C(T)}{T}  =  \gamma_1 + 2.88 \frac{m  \sqrt{2 m_z}}{\pi^2} \sqrt{T}  + \cO(T)
\eeq
The same behavior holds at a finite $\mu$, when $T > |\mu|$.
Observe that the prefactor for $\sqrt{T}$ term is positive, opposite to that of $T^2/\mu^{3/2}$ term. This implies that the temperature
dependence of $C(T)/T$ changes sign at some positive $\mu$.
The actual $T$ dependence of $C(T)/T$, obtained without expanding in $T$, is presented in Fig. \ref{fig:CvFree_1}a, and $\gamma_1$ and
$\gamma_3$ extracted
from fitting this $C(T)/T$ by $\gamma_1 + \gamma_3 T^2$
in different windows of $T$ are shown in Fig.\ref{fig:CvFree_2}b,c and in Ref.\onlinecite{SI}.
We see that
 both $\gamma_1$ and $\gamma_3$
 depend on where the $T$ window is set, and $\gamma_3$ as a function of doping changes sign at some $x >
x_c$, i.e.,
at some positive $\mu$, as expected.

{\it Interaction contribution to $C(T)$.}~~~At a qualitative level, the free-fermion formula for $C(T)$ is consistent with the data. At
the quantitative level, it strongly differs from the measured $C(T)$, even if we would use a renormalized dispersion with larger
effective density of states. To see the inconsistency, we compare in Fig.\ref{fig:CvFree_2}b,c the theoretical and experimental doping
dependence
of $C(T)$ and particularly the values of $\gamma_1$ and $\gamma_3$ fitted over various temperature ranges. We see that the
magnitude of $C(T)/T$ for free fermions and the strength of doping variation of $\gamma_3$, extracted from it, is much smaller than in
the data. These discrepancies call for the analysis of interaction contributions to $C(T)$.

A fully renormalized fermion-fermion interaction can be decomposed into effective interactions in the charge and in the spin channel. For
systems with
screened Coulomb repulsion, the effective interaction in the spin channel get enhanced and, if the system is reasonably close to a Stoner
instability, can
be viewed as mediated by spin fluctuations. Na$_x$CoO$_2$ does develop a magnetic order at $x > 0.75$
\cite{Johannes,Helme,Motohashi,Bayrakci}, and it
seems reasonable to expect that magnetic fluctuations develop already at $x \approx x_c$.

The spin-fluctuation contribution to the thermodynamic potential is given by~\cite{millis,de,lm}
\beq
\Omega = \Omega_0 + \int \frac{d \omega}{\pi} n_B(\omega) \int \frac{d^3 q}{(2 \pi)^3} \Im \ln \chi^{-1} (q, \omega)
\label{1}
\eeq
where $\Omega_0$ is the free-fermion part, $n_B$ is the Bose function, and $\chi (q, \omega)$ is fully renormalized dynamical spin
susceptibility.

To obtain $\chi (q, \omega)$ we use the same strategy as in earlier works~\cite{acs,betouras}:
compute first the static spin susceptibility $\chi_0 (q, \omega =0)$ of free fermions, then collect RPA-type  renormalization and convert
$\chi_0 (q, \omega=0)$ into full static $\chi (q, \omega=0)$, and
 then compute the bosonic self-energy coming from the interaction with
low-energy fermions and obtain the full dynamical $\chi (q, \omega)$ at low frequencies.  The result is\cite{SI}
\beq
\chi^{-1}(q,\omega) = \frac{{\bar \chi}}{\xi^{-2} + (q-k_{F1})^2 +b \omega^2 - i \gamma \omega}
\label{chi}
\eeq
where $\xi$ is a magnetic correlation length
 and the last term is the Landau damping.
  The sign of $\gamma_3$ term in $C(T)$ depends on the sign of $b$ -- the  prefactor  for the $\omega^2$  term
   (see Eq. (\ref{fr_aa}) below).
 To obtain $b$ in our case we first evaluated the susceptibility of free fermions $\chi_0 (q, \omega)$ and then obtained $\chi (q, 
\omega)$ using RPA.
For most relevant  $q \approx k_{F_{1}}$ we obtained
  (see [\onlinecite{SI}] for details)
\beqa
\label{Susceptibility_1}
 \chi_0 (q,\omega) &=& \frac{\sqrt{m m_z}}{4 \pi^2 v_{F1}} \left [\left(\omega -\tilde\mu \right) \log \left(|\omega
-\tilde\mu | \right)+ \right. \nn \\
&& \left. -\left(\omega +\tilde\mu \right) \log \left(|\omega +\tilde\mu |\right)\right] + ...
\eeqa
where $\tilde\mu = \mu -  (q- k_{F_1})^2/(2m)$ and
dots stand for  regular terms.
 Expanding in $\omega$ and substituting into RPA formula we obtain
 \beq
\label{eq b}
b =  \frac{\sqrt{m m_z}}{4 \pi^2 m_z a_0 v_{F1}} \frac{1}{\tilde \mu}
\eeq
\begin{equation}
\gamma=
 \frac{\sqrt{m m_z}}{4 \pi m_z a_0 v_{F1}} \theta(\tilde\mu)
+
\frac{1}{\sqrt{3} \pi  v_{F1}^2 m_z a_0 a_z },
 \label{eq gamma}
\end{equation}
 where $a_0$ is of order of lattice spacing in $xy$ plane, $a_z$ is inter-layer spacing. Note that near $\mu=0$ the quadratic coefficient
$b$ is singular and its dependence on $q$ becomes important.
 The $1/{\tilde \mu}$  dependence of  $b$ originates from the singularity in the derivative of density of states at the Lifshitz 
transition.
The $T^3$ term in $C(T)$ at $x < x_c$ and small $T$ ( $T < |\mu|$ and
 $T< 1/(\xi^2 \gamma)$ )
 comes from expanding $\Im
\ln \chi^{-1}$ in
(\ref{1}) to order $\omega^3$ and integrating over $q$ near $q=k_{F1}$. When $|\mu|>\xi^{-2}/m$ the $q$-dependence of $b$ and $\gamma$
may be neglected and we obtain
 \beq
\gamma_3=  \gamma k_{F1} \xi^3 \frac{\pi^3}{10} \left(-4 b - (\gamma \xi)^2\right)
 \label{fr_aa}
\eeq
The \eq{eq b} for $b$ suggests a singular behavior of $\gamma_3$ near $\mu=0$.
For small $|\mu| < \xi^{-2}/m $ the singularity is smoothed by $q$-dependence of
$\gamma$ and $b$ and \eq{fr_aa} needs to be replaced by the result of numerical integration. The results,
 in particular a sharp maximum in $\gamma_3$ near $x_c$,
 are in good
agreement with experiment, see Fig.\ref{fig:CvFree_2}b,c.

At higher temperatures, when $T > 1/(\xi^2 \gamma)$
 we find  that the system enters into a quantum-critical regime
 where the system behavior is the same as at $\xi^{-1} =0$.
 The form of $C(T)/T$ at such temperatures
 in principle  depends on the
 effective dimensionality of spin fluctuations around $q=q_0$ (see Ref. \cite{SI}).
 We find, however, that
 such dimension-specific
behavior holds only at high $T$, while in the intermediate
regime $T \gtrsim 1/(\xi^2 \gamma)$,  $C(T)/T$ can be well fitted by $\log T$  even for
effectively 1D spin fluctuations.
This agrees with the data which show a $\log T$ behavior even at doping $x=0.65$, see Fig. \ref{fig:CvFree_3}.
As $\xi$
and $\gamma$
increase at larger $x$,
 the lower boundary of $\log T$ behavior of $C(T)/T$
 stretches to progressively smaller
$T$
and a prefactor of $\log T$ grows,
in agreement with
 the experiments at higher doping
  (Ref. [\onlinecite{Balicas,Okamoto}]).

For quantitative comparison with the data we compute the dynamical part of particle-hole bubble without
expanding in frequency and use (\ref{1}) to compute the thermodynamic potential and the specific heat.
To estimate $\xi$ we use the experimental data for $\chi (0,0)/\gamma_1$ at $x \approx x_c$ and our numerical RPA
result for the prefactor for $(q-q_0)^2$ term in $\chi^{-1} (q, \omega)$. Extracting $\xi$ from these data we obtain
$\xi \approx 7 a_0$ near $x=0.62$ and it grows with the doping.
For better comparison
 we subtract from the data the contribution from phonons $C_{ph} \approx T^3 \cdot 0.07 mJ K^{-4} mol^{-1}$, which only weakly depends on
doping \cite{comm3}.
The results are shown in Fig. \ref{fig:CvFree_1}b and Fig. \ref{fig:CvFree_2}b,c. We see that theoretical and experimental $C(T)$ agree 
quite well over a wide range of
temperatures,
and the agreement between $\gamma_1$ and $\gamma_3$, extracted from the data and from spin-fluctuation theory, is also very good.
We emphasize that the doping variation of $\gamma_3$ is not affected by the phonon contribution and thus measures solely the contribution
to $C(T)$ from
spin fluctuations. From this perspective, a good agreement with the data is an indication that  magnetic fluctuations with large
$q=k_{F1}$ are strong in
Na$_x$CoO$_2$ near the LTT. The $\log T$ behavior of $C(T)/T$, which we found at $T \sim 3 - 10 K$ for $x \approx 0.7$
is also consistent with the data, see Figs.\ref{fig:CvFree_3}. Finally, we note that the experimental data on $\gamma_1$, fitted at $T
\sim 10$K,
show a small discontinuity as a function of doping, Figs.\ref{fig:CvFree_2}b,c,  which is expected if the LTT is first order
~\cite{Slizovskiy}. The jump in $\mu$ is estimated to be 5 to 10 meV. When we take this into account, we
obtain
a sharper doping dependence of $\gamma_3$, leading to an even better agreement with the data.

{\it  Conclusions.}~~~~ In this work we have analyzed the specific heat in the layered cobaltate  Na$_x$CoO$_2$.
 Near $x=0.62$ the system exhibits
  a non-analytic temperature dependence and strong doping variation of the specific heat coefficient $C(T)/T$.
We explained the data based on the idea that at $x_c =0.62$ the system undergoes a LTT in which  a new electron pocket appears.
We demonstrated that the non-analytic temperature dependence of $C(T)/T$ at $x=x_c$ and its strong doping variation is quantitatively
reproduced
if
interaction is mediated by spin fluctuations peaked at the wave-vector which connects the original and the emerging Fermi
surfaces.
 We argued that
 the observed
 $\log T$ behavior of $C(T)/T$ at
 larger dopings
$0.65\lesssim x<0.75$
 is an indication that
  the system enters into the magnetic critical regime.

We acknowledge useful discussions with S. Carr, A. Katanin,  F. Kusmartsev, D. Maslov, J. Quintanilla, S. Shastry, J.
Zaanen. We thank Y. Okamoto and Z. Hiroi for communication and for sending us the experimental data. The work was supported by the EPSRC
grants EP/H049797/1 and EP/l02669X/1 and (J.B and S. S) and by the DOE grant DE-FG02-ER46900 and a Leverhulme Trust visiting professorship
held at Loughborough University (A.V.C.).

\section*{Supplementary}
\subsection{Magnetic susceptibility.}
We follow earlier works~\cite{acs} assuming that the static magnetic susceptibility and regular part of its frequency  dependence
are governed by high-energy processes. 
Then, the Landau damping term and the singular part of the remaining frequency dependence 
come from fermions with low energies and can be obtained within the low-energy spin-fermion model. As a result, the contributions from 
high-energy fermions can be incorporated into the static susceptibility through a tunable ``magnetic correlation length'' parameter, and 
we then focus on the particle-hole contributions coming from low fermion energies.

The free-fermion susceptibility $\chi_1$, coming solely from  the large cylindrical Fermi-surface (FS) is  a 2D Lindhard function:
\beqa \label{Lindhard}
 \Im \chi_1(\omega) = \omega \frac{2 \chi_0}{\sqrt{3} k_{F1} v_{F1} a_z } = \omega \frac{1}{\pi \sqrt{3} v_{F1}^2 a_z} \\
 \Re \chi_1(\omega) \approx \frac{k_{F1}}{2 \pi v_{F1} a_z}
\eeqa
where $a_z$ is inter-layer lattice spacing.
This $\chi_1$ does not carry any interesting frequency or chemical potential
dependence.

On the contrary, the susceptibility
$\chi_{12} (q, \omega)$, coming from the particle-hole processes with total momentum $q \approx k_{F1}$, depends non-trivially on $\omega$ 
and $\mu$.
We take the momenta $q$ to be near the distance between Fermi momentum for the hole FS and $\Gamma$ point where electron FS emerges for $x 
 > x_c$, i.e., consider
 $q =k_{F1} + \tilde q$ and assume $\tilde q$ to be small.
Because $k_{F2}$ for the electron pocket is either zero ($x<x_c$) or very small ($x > x_c$), we deal with a special case when the 
frequency may exceed the Fermi energy of the small pocket. This gives rise  to a non-linear frequency dependence of the imaginary part of 
the susceptibility at $q \approx k_{F1}$.
To simplify the discussion, we approximate the hole dispersion as purely 2D and the dispersion near $\Gamma$ as a 3D parabola.
Evaluating the imaginary part of the particle-hole bubble involving hole-like and electron-like excitations, we obtain
\beq
  \Im \chi_{12}(q,\omega) = \frac{1}{16 \pi^2 v_{F1}} S
\eeq
Here
 $S$ is the area in the
$k_y , k_z$ plane,
 where
  $\mu-\omega < \tilde q^2/(2 m) + k_y^2/(2 m) + k_z^2/(2 m_z) < \mu + \omega$.
This area is a ring for
  $|\omega| < \mu - \tilde q^2/(2 m) \equiv \tilde \mu $ and an ellipse otherwise;  the ellipse shrinks to
an empty set if $\tilde \mu + |\omega|<0$.  In explicit form:
\beqa
 S &=& 2 \pi  \sqrt{m m_z} \( ( \tilde \mu  + \omega) \theta(\tilde \mu + \omega) - \right.\nn \\
&& \left. - (\tilde \mu - \omega) \theta(\tilde\mu -
\omega)\)
\eeqa
where $\tilde \mu = \mu - \tilde q^2/(2m)$. We also need an extra factor of 2, if we add the contribution from the opposite patch of the 
large pocket.

\begin{figure}
\includegraphics[scale=0.55]{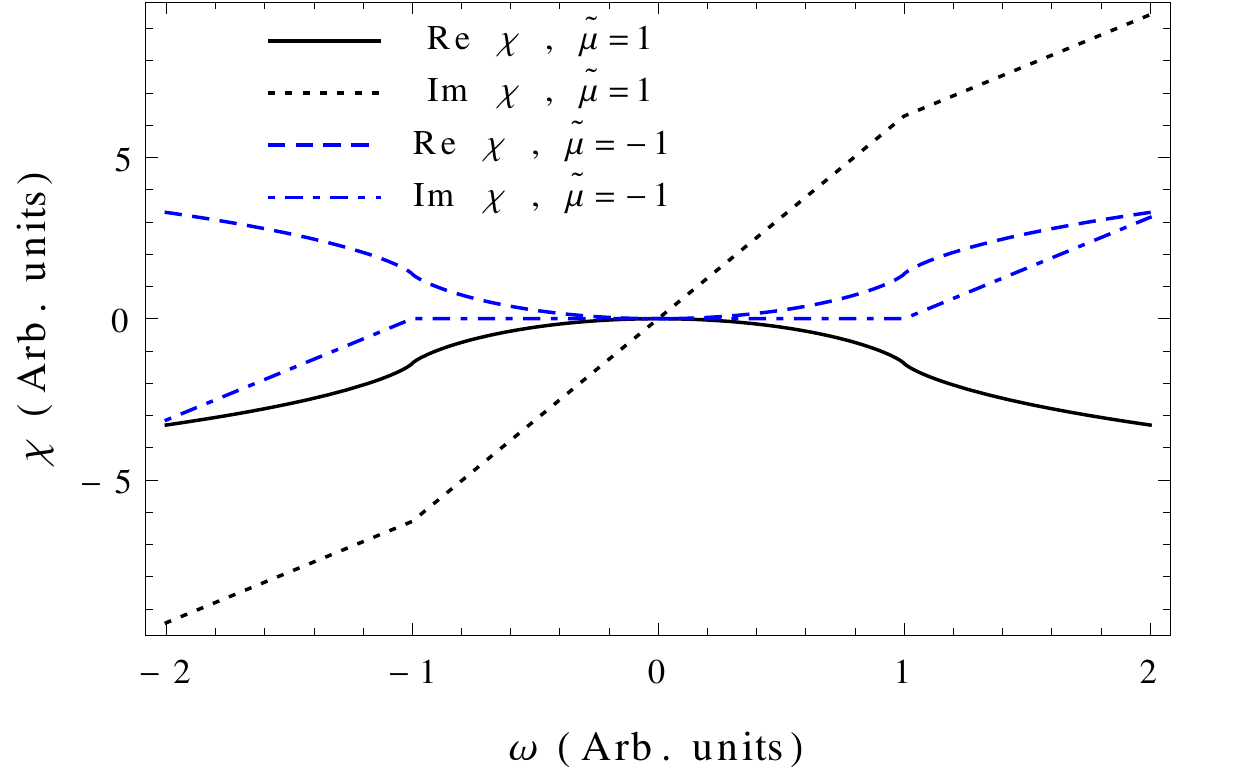}
\caption{\label{fig:ChiSketch}
The frequency dependence of the susceptibility $\chi_{12}(\omega)$ }
\end{figure}

Analyzing $S(\omega, \tilde q = 0)$, we find that it has a linear frequency dependence at the lowest frequencies at $\mu
>0$, when the small pocket is present, then there is a cusp at $\omega = \mu$,  and then another linear dependence, with twice smaller
slope.
  For $\mu<0$, when there is no pocket but the dispersion has a local minimum at $\Gamma$, the slope is zero at
  $\omega < -\mu$ and becomes finite only after  the cusp at $-\mu$,
   see Fig.\ref{fig:ChiSketch}. At a nonzero $\tilde q$ the results are the same as at $\tilde q =0$ if one replaces $\mu$ by
$\tilde \mu$.

The frequency-dependent part of $\Re \chi_{12} (q, \omega)$ can be computed elegantly from
Kramers-Kronig transformation. As the second frequency derivative of the imaginary part is a delta-function at the cusp, it is easy to 
compute the second frequency derivative of the real part:
\beq \label{Real Part}
  \d_\omega^2 \Re \chi_{12}(\omega) = \frac{\sqrt{m m_z}}{4 \pi^2 v_{F1}} \(\frac1{\omega - \tilde\mu} - \frac1{\omega + \tilde\mu} \)
\eeq
This expression is singular  at frequencies $\omega = \pm \tilde \mu$,
see Fig. \ref{fig:ChiSketch}.
It is essential for our analysis that $ \d_\omega^2 \Re \chi_N(\omega)>0$ for $\tilde\mu<0$ and that it diverges when $\tilde \mu \to 0$.

Integrating Eq. (\ref{Real Part}) over $\omega$ we obtain
the full analytic expression for the frequency dependence of
susceptibility:
\beqa \label{Susceptibility}
 \chi_{12}(q,\omega) &=& \chi_{12}(q,0) +\frac{\sqrt{m m_z}}{4 \pi^2 v_{F1}} \left [\left(\omega - \tilde\mu \right) \log \left(\omega
-\tilde\mu \right)+ \right. \nn \\
&& \left. + \tilde\mu
\log \left(\tilde\mu^2 \right)  -\left(\omega +\tilde\mu\right) \log \left(-\omega - \tilde\mu\right)\right]
\eeqa
The result is presented in
 Fig. \ref{fig:ChiSketch}. At $\tilde q = \pm k_{F2}$,  $\tilde \mu = 0$, and the singularity in $\chi_{12}(\omega)$ is located at zero 
frequency. The static part
\beq
\chi_{12}(q,0) \approx a_0 m_z (q-q_0)^2 + const
\eeq
 has non-universal high energy contributions
 which have to be computed
numerically and are effectively included in our calculations through the value of the correlation length. The parameter
 $a_0$ is of the order of the lattice spacing
 in $xy$ plane.

\subsection{Direct derivation of the prefactor of the $\omega^2$ term in spin susceptibility at $q=k_{F1}$.}
In this section, we offer an alternative derivation of the prefactor $b$. Consider the case $\mu<0$ and $q=k_{F1}$ along the $x$-axis. 
Working in Matsubara formalism we obtain for the free-theory susceptibility
at $T=0$:
 \beq
 \chi_0(k_{F1},i\Omega) = - \int \frac{d^3 k\, d\omega}{(2 \pi)^4} G(k,i\omega) G(k+k_{F1}, i\omega+i\Omega)
\eeq
The Green's functions are
\beq
G(k+k_{F1},i \omega+i\Omega) =  \frac{1}{i (\omega + \Omega) - \left(\frac{k_x^2 + k_y^2}{2 m} + \frac{k_z^2}{2 m_z} + |\mu|\)}
\eeq
\beq
G(k,i\omega) = \frac{1}{i \omega + v_{F1} k_x}
\eeq
One Green's function is taken near the $k=0$ (where the small pocket
is due to appear) and another is near the large hole pocket. The contribution is doubled to account for the reverse situation. Expanding
in $\Omega$ we get the quadratic term:
\beqa
 \chi_0^{(2)}(k_{F1},i \Omega) = + 2 \Omega^2 \int \frac{d k_z\, dk_x\, dk_y \, d\omega}{(2 \pi)^4} \nn
\\ \frac{1}{\left(i \omega -
\left(\frac{k_x^2 + k_y^2}{2 m} + \frac{k_z^2}{2 m_z} + |\mu|\)\)^3 (i \omega + v_{F1} k_x) }
\eeqa
  For $k_x>0$ the frequency integration contour passes between the poles and we get
\beqa
 \chi_0^{(2)}(k_{F1},i \Omega) = -2 \Omega^2 \int \frac{d k_z \, dk_y}{(2 \pi)^3} \int_0^\infty dk_x
\nn \\ \frac{1}{\left(v_{F1} k_x + \frac{k_x^2 + k_y^2}{2 m} + \frac{k_z^2}{2
m_z} + |\mu|\)^3 }
\eeqa
for small $|\mu|$ the integration is peaked at small $k$ and we can neglect $k_x^2/(2 m)$ term compared to $v_{F1} k_x$.
Then defining $x= k_x v_{F1}/|\mu|$, $y=k_y/\sqrt{2 |\mu| m}$ and $z=k_z/\sqrt{2 |\mu| m_z}$ we get:
\beqa
 \chi_0^{(2)}(k_{F1},i\Omega) &=& -2 \Omega^2 \frac{2\sqrt{m m_z}}{v_{F1} |\mu|} \int_{-\infty}^{\infty} \frac{d z \, dy}{(2 \pi)^3}
\int_0^\infty dx \cdot
\nn \\ &&\cdot \frac{1}{\left(x + y^2 + z^2 + 1\)^3} \nn\\
&=& - \Omega^2 \frac{\sqrt{m m_z}}{4 \pi^2 v_{F1} |\mu|}
\eeqa
  The continuation to real frequencies provides a positive coefficient of $\Omega^2$ in $\chi_0$, which behaves as $1/|\mu|$ for $\mu<0$.
This coincides with the more general \eq{Real Part}, if we set $\omega = 0$ and $q=k_{F1}$ in \eq{Real Part}.
This unconventional result appears because the momentum $q$ connects a Fermi-surface with a region where all states are above the Fermi
level.
When the small pocket appears and fermions are connected by the vector $q$ are simultaneously at the Fermi-surface, an opposite sign is
obtained.

\subsection{Temperature expansion of the specific heat.}
At low temperatures, when $|\omega| < |\mu|$ and $T<\mu$, we can expand the full free-particle susceptibility
$\chi_0 (q, \omega) = \chi_1 (q, \omega) + \chi_{12} (q, \omega)$ in frequency as
\beq
\chi_0(q, \omega) = \chi_0 - m_z a_0 (q-q_0)^2 +b_0 \omega^2 + i \gamma_0 \omega
\eeq
From \eqs{Lindhard}{Susceptibility} we extract:
\beqa
 \gamma_0 =
\frac{\sqrt{m m_z}}{2 \pi v_{F1}} \theta(\tilde\mu)
+ \frac{1}{\sqrt{3} \pi  v_{F1}^2 a_z }
 \label{gamma0} \\
b_0 =  - \frac{\sqrt{m m_z}}{4 \pi^2 v_{F1} \tilde \mu}
\eeqa
The full RPA magnetic susceptibility
$\chi (q, \omega) = \chi_0 (q, \omega)/(1-U \chi_0 (q, \omega)$ is then expressed as
\beq
\chi (q,\omega) = \frac{{\bar \chi}}{\xi^{-2} + (q-q_0)^2 + b \omega^2 - i \gamma \omega}
\eeq
where
${\bar \chi} \approx \chi_0/(U m_z a_0)$, $\xi^{-2} = (1/U-\chi_0)/(m_z a_0)$, and
\beq
\label{eq b suppl}
b = - \frac{\chi_0 b_0 + \gamma_0^2}{\chi_0 a_0}  \approx - \frac{b_0}{m_z a_0} = \frac{\sqrt{m m_z}}{4 \pi^2 m_z a_0 v_{F1}} \frac{1}{
\tilde \mu}
\eeq

\beq \label{eq gamma suppl}
\gamma=\frac{\gamma_0}{m_z a_0} = \frac{\sqrt{m m_z}}{2 \pi m_z a_0 v_{F1}} \theta(\tilde\mu)
+ \frac{1}{ \sqrt{3} \pi  v_{F1}^2 m_z a_0 a_z }
\eeq

The spin-fluctuation contribution to the grand thermodynamic potential is~\cite{millis}:
\beq
\Omega_{int} = \int \frac{d \omega}{\pi} n_B(\omega,T) \int \frac{d^3 q}{(2 \pi)^3} \Im \ln \chi^{-1}
\eeq
$n_B(\omega,T)$, however the form of temperature dependence of $\Omega_{int}$ depends on the frequency dependence of $\chi (q, \omega)$.  
Expanding the integrand in frequency and differentiating  $\Omega_{int}$ over $T$ we obtain the temperature expansion of the interaction 
contribution to the entropy $S =-\d
\Omega_{int}/\d T$.
\beqa
  &S = \frac{T}{3 (2 \pi)^{D-1}} \int \frac{\gamma \, d^D
q}{\xi^{-2} + (q-q_0)^2} -  \\
& - \frac{T^3}{15 (2 \pi)^{D-3}} \(  \int
\frac{ \gamma  b \, d^D
q}{\left(\xi^{-2} + (q-q_0)^2\)^2} + \frac13 \int\frac{d^D q \gamma^3}{\left(\xi^{-2} + (q-q_0)^2\)^3} \) \nn
\eeqa
The momentum integral is  peaked at $q=q_0 \approx k_{F1}$ and
 we assume that it is cylindrically symmetric (the actual dispersion suggests that
$q_z = \pi$ may be more important than other values of $q_z$,  but this only changes the overall prefactor).
Extracting $C(T)$ from the entropy we
 obtain
 $C(T)= T \gamma_1 + T^3 \gamma_3$, where, if we neglect $q$-dependence of $\gamma$ and $b$,
\beq \label{gamma_3}
\gamma_3=  \gamma k_{F1} \xi^3 \frac{\pi^3}{10} \left(-4 b - (\gamma \xi)^2\right)
\eeq
These are the expressions used in the main text.
If we are close to $\mu=0$, then $b$ is singular, so we cannot neglect its $q$ dependence. The integration has been done numerically,
while the qualitative analytic picture is given below.
At small enough $\mu$,
 $1/\tilde\mu$ has to be replaced by
 $-(2m)/(q-k_{F1})^2$.  Typical $q-k_{F1}$ is of order $\xi^{-1}$,
hence $b$ saturates at the value of
 order $\xi^2 (m^3 m_z)^{1/2}/(m_z a_0 v_{F1})$.
Note that the ratio $4|b|/(\gamma \xi)^2$ does not depend on $\xi$.
This ratio exceeds one, for the dispersion we consider, hence, according to \eq{gamma_3},
$\gamma_3 >0$. A positive $\gamma_3$, which increases as $x$ approaches $x_c$ from below, is precisely what the data show
(see Figs.\ref{extra} c,d).

At $\mu >0$ ($x \geq x_c$), when the new pocket appears,
 $\gamma$
 changes by a finite amount and
 $b$  evolves from a negative constant at $\mu = 0$ to a positive $b \propto 1/\mu$ given by
 (\ref{eq b suppl}) for $\mu > \xi^{-2}/(2 m)$.  The positive $b$ is consistent with earlier result~\cite{millis} for a
  single 3D FS as in both cases $q \approx k_{F1}$ connects points on the FS and our small pocket is three-dimensional.
As a result,
$\gamma_3$  rapidly decreases as $x$ increases above $x_c$,
  changes sign and becomes negative.
This is again consistent with the data.

\subsection{\it Specific heat of free fermions for $|\mu| \ll T$.}
Let us fix $x=x_c$, so that $\mu(T=0) = 0$.   The 3D pocket produces
a singularity in the density of states:
\beq
\rho = B + A \sqrt{\mu}
\theta(\mu)
\eeq where in our case $B = \frac{k_{F1}}{\pi v_{F1}}$ and $A = \frac{m  \sqrt{2 m_z}}{\pi^2}$.
The grand canonical potential is $\Omega = -\int \tilde \rho(e) n_F((e-\mu)/T) de $,  where $\tilde \rho(\eps) = \int^\eps \rho(e) de $.
Then the entropy is $S = -T^{-2} \int \tilde \rho(e) n_F'((e-\mu(T))/T) (e - \mu)  de =  -\int \tilde \rho(e T) n_F'(e-\mu/T) (e-\mu/T)
de $.
The specific heat is obtained by substituting $\mu(T)$ and evaluating $C/T = \left( \frac{\d S}{\d T}\right)_N$.

The condition on the chemical potential $\mu(T)$ is
\beq
 \int ( n_F\left(\frac{e- \mu(T)}{T}\right) - \theta(-e) )  (B + A e^{1/2} \theta(e)) de  = 0
\eeq
this equation can be expanded is series in $\mu(T)/T$:
\beq
 \mu(T) = -\frac{0.678\,  A \, T^{3/2}}{0.536 \, A \, \sqrt{T}+B} + \cO((\mu/T)^2)
\eeq
This expansion is valid for moderate temperatures, where $A \sqrt{T} \ll  B$, then, indeed $\mu/T \ll 1 $, the actual expansion parameter
is $A \sqrt T / B $.

For the specific heat we obtain
\beq
C/T =
\frac{\pi^2}{3}  B + 2.88 A \sqrt{T} -\frac{0.88 A^2 T}{B}  + B \cO \(\frac{A \sqrt T}{B}\)^3
\eeq

\subsection{Quantum criticality near the transition to ordered phase.}
If the small pocket and the large pocket have parts with matching curvatures, the susceptibility will be strongly peaked
at the momentum vector connecting them, which may result in spin density wave magnetic order.  This is indeed what happens
when the small pocket has grown sufficiently large at $x>0.75$ \cite{Korshunov, Kuroki2007}.
For
smaller doping, the nesting is not good enough for magnetic order, but magnetic correlation length
$\xi$ is
 nevertheless
 large.    When $T \ll \eps_\xi$ and $T \ll \mu$, a
  regular
  Fermi-liquid
expansion of $C(T)/T$ in powers of $T^2$
 works.  At slightly higher temperatures, when $\xi^{-2}$ cannot be assumed large (see \eq{CriticalCondition})
  this expansion does not hold.
 This temperature regime is relevant to the description of
  the behavior of $C(T)/T$ at intermediate $T$ at $x$ near $x = x_c =0.62$ and down to quite low $T \sim 0.1 K$ at
 $x = 0.747$ (Ref. \onlinecite{Balicas}),
  which is very close to $x=0.75$ at which $\xi=\infty$.
   The fact that critical
behavior extends to such low temperatures is remarkable.

  \begin{figure*}[htp]
$\begin{array}{cc}
\hspace{-0.8 cm}\includegraphics[scale=0.49]{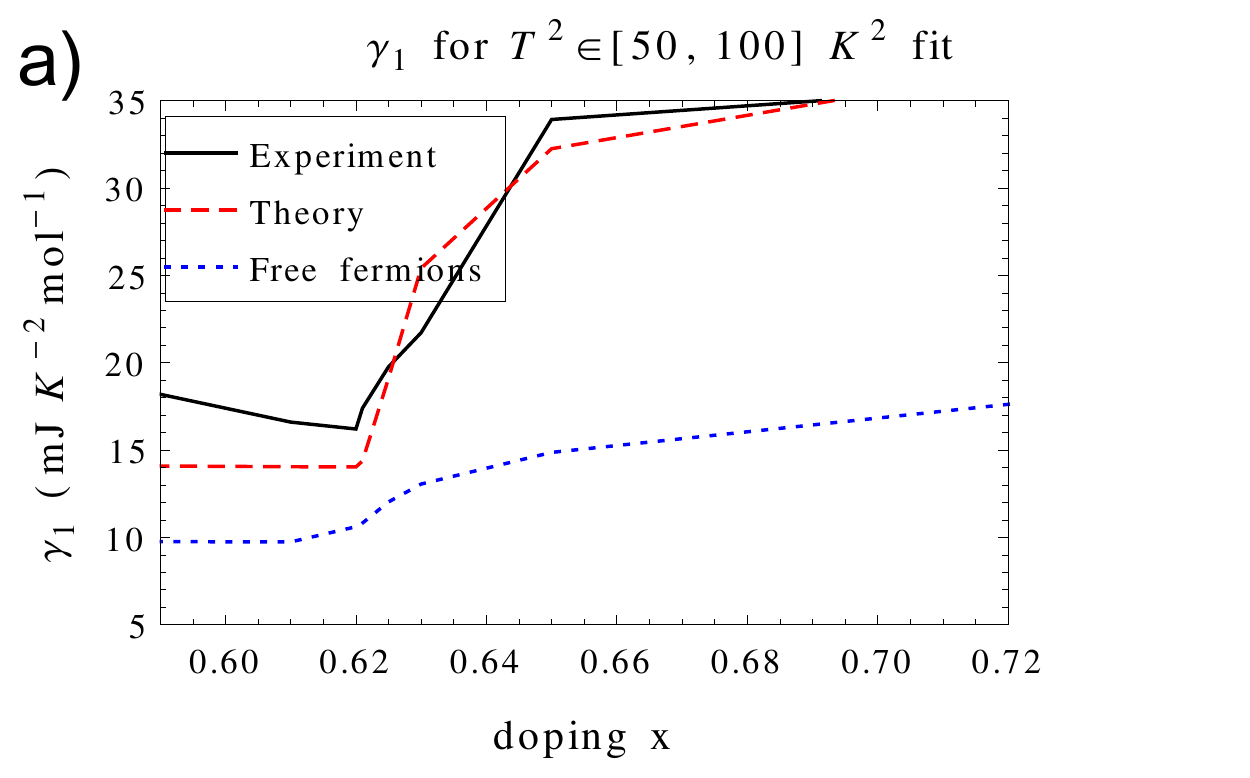} \hspace{-1cm}&
\hspace{-0.3 cm} \includegraphics[scale=0.49]{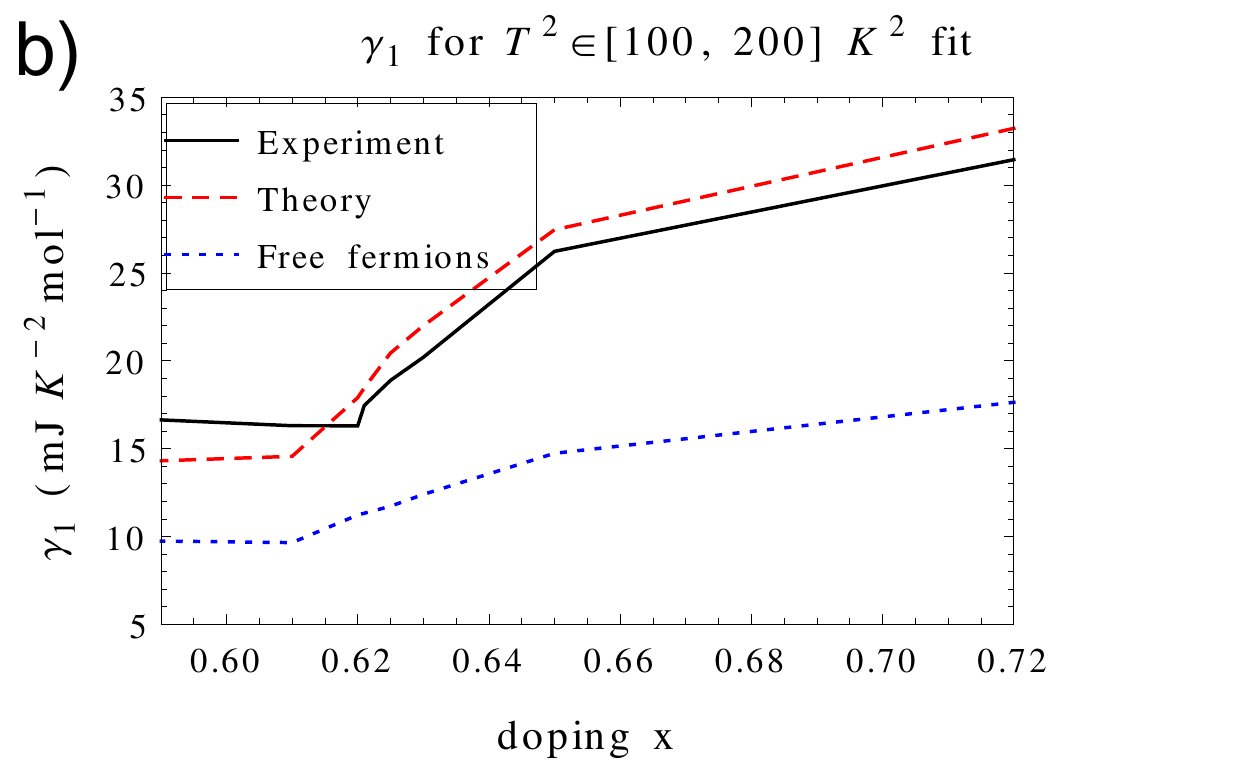} \\
\hspace{-0.5cm}  \includegraphics[scale=0.49]{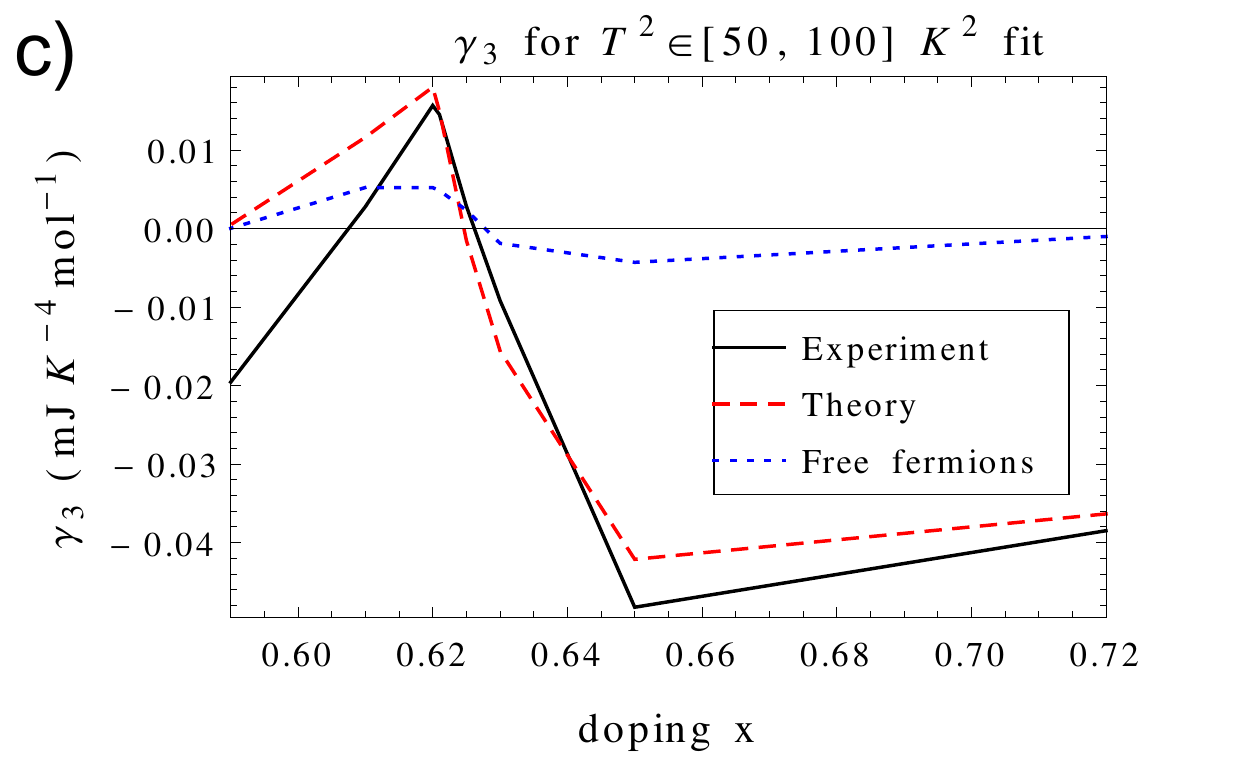}&
\hspace{-1cm}\includegraphics[scale=0.49]{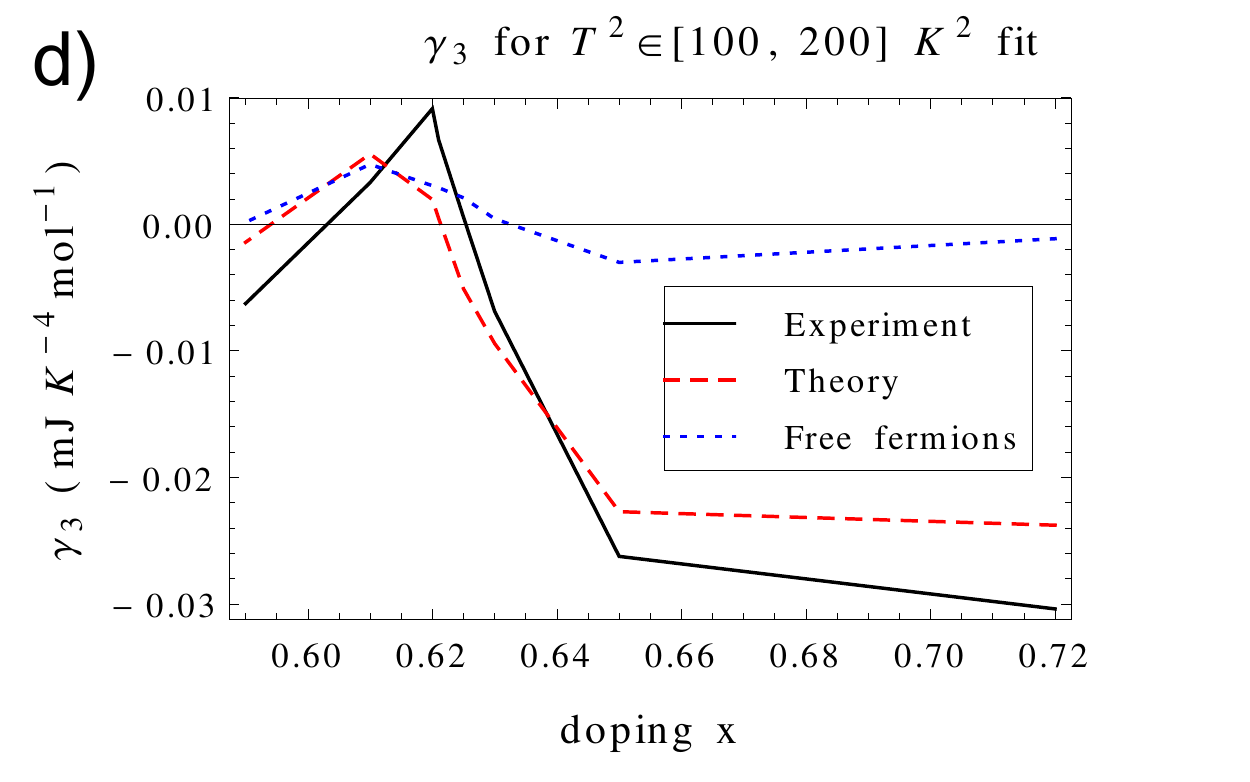}
\end{array}$
\vspace{-0.5cm}
 \caption{\label{extra}
 The fits of the experimental data for the specific heat $C(T)/T$ for various $x$ and
 theoretical formulas for free fermions  and for fermions with magnetically-mediated interaction with $\xi=7$
 to $C(T)/T$ to $C(T)/T = \gamma_1 + \gamma_3 T^2$ in two temperature ranges centered at different $T$.}
\end{figure*}

For analytic estimate, we set
$\omega$ to be of order $T$ and using earlier results obtain
\beqa\label{CriticalEstimate1}
  &C(T) \sim \\ \nn & \left. - \int d^3 \vec q  \,\Im \left(\log \left (\xi^{-2} + \left(\sqrt{q_x^2+q_y^2}-k_{F1}\)^2 - i \gamma \omega 
\) \) \right|_{\omega
\approx T}
\eeqa
where $\gamma$ is given in \eq{eq gamma suppl}.  Here we integrate over cylindrical shell of momenta near the vectors that connect
$\Gamma$-point  with large pocket.  Here we assumed the large pocket to be cylindrical, but this assumption can be easily relaxed
not changing the result.  There is a whole 2D surface of important momenta that contribute to near-critical
spin fluctuations (kinetic part $(\sqrt{q_x^2+q_y^2}-k_{F1})^2$ is degenerate in $q_z$ and angle).
When $T \gtrsim \xi^{-2}/\gamma$, the system 
 enters the critical region.  Deep in
 this regime,
the specific heat behaves as
\beq
  \frac{C(T)}{T} \sim \frac{1}{T} \Im \sqrt{\xi^{-2} + i T \gamma}  \to   \frac{\gamma}{\sqrt{T}}
\eeq
This behavior can be traced back to the fact that the dispersion of spin fluctuations near $q=q_0$
is effectively one-dimensional since it is independent on the direction
of ${\bf q}$  in the $xy$ plane and on $q_z$. 
For effectively $2D$ dispersion, we would obtain $C(T)/T \propto \log T$, and for $3D$
dispersion $C(T)/T$ would remain finite.

We have found, however, that in our, effectively one-dimensional, case the $1/\sqrt{T}$ dependence of $C(T)/T$ holds only at high $T$,
while in a wide range of temperatures
the function $\Im(\sqrt{a + i T})/T $ can be well approximated numerically by  $(0.44 - 0.095 \log\frac{T}{a})/\sqrt{a}$.
This behavior holds at
$a \lesssim T \lesssim 40 a$.
In our case $a=\xi^{-2}/\gamma$
and we expect a $\log T$ behavior at $T \gtrsim \xi^{-2}/\gamma$.  Assuming that the largest contribution to $\gamma$ comes from the
approximate nesting of small and large Fermi-surfaces, we get $\gamma^{-1} \sim a_0 v_{F1}$, so
\beq \label{CriticalCondition}
  T \gtrsim \frac{a_0 v_{F1}}{\xi^2} = \frac{v_{F1}}{a_0} \left(\frac{a_0}{\xi}\)^2
\eeq
the factor $v_{F1}/a_0$ can be estimated to be of the order of effective hopping $t_1 \approx 1000$K. Hence, the correlation length of
order of 10 lattice spacings can bring the $\log T$ behavior to the range of several Kelvins.

The calculation can be easily repeated if one assumes that lattice effects produce a significant lifting of degeneracy of the
kinetic term for spin fluctuations. E.g. let us replace $ (\sqrt{q_x^2+q_y^2}-k_{F1})^2 \to ((q_x-k_{F1})^2 + q_y^2)$, then the
integration in \eq{CriticalEstimate1} gives logarithmic behavior of $C/T$  in the same range $T >  \xi^{-2}/\gamma$.
In the main text we used the approximate theoretical formula $C(T)/T \propto \log T$ to fit the experimental data.

\subsection{Fits of $C(T)/T$ to $\gamma_1 + \gamma_3 T^2$ in temperature intervals centered at different $T$. }
As emphasized in the main text, the T-dependence of the experimental $C(T)/T$ around $x =0.62$ is stronger than $T^2$, 
as evidenced by the fact that the fits of the data on $C(T)/T$  to
$\gamma_1 + \gamma_3 T^2$ behavior~ in finite intervals around different $T$ yield larger $\gamma_3$ with decreasing $T$. We show the fits 
in two temperature intervals centered at different $T$ in Fig. \ref{extra}.

\end{document}